\documentclass[a4paper,twocolumn,11pt,accepted=2021-06-24]{quantumarticle}
\pdfoutput=1

\usepackage{datetime}
\usepackage{graphicx,epsfig,float}
\usepackage{color}
\usepackage[usenames,dvipsnames]{xcolor}
\usepackage{amsmath,bbm,amssymb,amsthm,array,hhline,tabularx,makecell}
\usepackage[colorlinks=true,linkcolor=MidnightBlue,citecolor=blue,filecolor=green,urlcolor=PineGreen]{hyperref}
\usepackage{dsfont} 
\usepackage{stmaryrd}
\usepackage{numprint}
\newcolumntype{Y}{>{\centering\arraybackslash}X}

\def\bb{\begin{eqnarray}}
\def\ee{\end{eqnarray}}
\newcommand{\ket}[1]{| #1 \rangle}
\newcommand{\bra}[1]{\langle #1 |}

\newcommand{\red}[1]{{\color[rgb]{1,0,0}{#1}}}

\newcommand{\blue}[1]{{\color[rgb]{0,0,1}{#1}}}

\usepackage{lineno}

\makeatletter
\newcommand*{\textoverline}[1]{$\overline{\hbox{#1}}\m@th$}
\makeatother

\newtheorem{property}{Property}
\newtheorem{condition}{Condition}
\usepackage{numprint}

\begin{document}

\title{Quantum erasing the memory of Wigner's friend}


\author{Cyril Elouard}
\email{cyril.elouard@gmail.com}
\affiliation{Department of Physics and Astronomy, University of Rochester, Rochester, NY 14627, USA}
\affiliation{QUANTIC lab, INRIA, 2 Rue Simone IFF, 75012 Paris, France}
\author{Philippe Lewalle}
\affiliation{Department of Physics and Astronomy, University of Rochester, Rochester, NY 14627, USA}
\author{Sreenath K. Manikandan}
\affiliation{Department of Physics and Astronomy, University of Rochester, Rochester, NY 14627, USA}
\author{Spencer Rogers}
\affiliation{Department of Physics and Astronomy, University of Rochester, Rochester, NY 14627, USA}
\author{Adam Frank}
\affiliation{Department of Physics and Astronomy, University of Rochester, Rochester, NY 14627, USA}
\author{Andrew N. Jordan}
\affiliation{Department of Physics and Astronomy, University of Rochester, Rochester, NY 14627, USA}
\affiliation{Institute for Quantum Studies, Chapman University, Orange, CA 92866, USA}

\newdate{subdate}{04}{07}{2021}
\date{subdate}

\begin{abstract}
The Wigner's friend paradox concerns one of the most puzzling concepts of quantum mechanics: the consistent description of multiple nested observers. Recently, a variation of Wigner's gedankenexperiment, introduced by Frauchiger and Renner, has lead to new debates about the self-consistency of quantum mechanics. At the core of the paradox lies the description of an observer and the object it measures as a closed system obeying the Schr\"odinger equation. We revisit this assumption to derive a necessary condition on a quantum system to behave as an observer. We then propose a simple single-photon interferometric setup implementing Frauchiger and Renner's scenario, and use the derived condition to shed a new light on the assumptions leading to their paradox. From our description, we argue that the three apparently incompatible properties used to question the consistency of quantum mechanics correspond to two logically distinct contexts: either one assumes that Wigner has full control over his friends' lab, or conversely that some parts of the labs remain unaffected by Wigner's subsequent measurements.  The first context may be seen as the quantum erasure of the memory of Wigner's friend. We further show these properties are associated with observables which do not commute, and therefore cannot take well-defined values simultaneously. Consequently, the three contradictory properties never hold simultaneously.
\end{abstract}

\maketitle

\section{Introduction}


In his famous gedankenexperiment, Wigner analyzes a setup in which a ``super-observer'' is assumed to be able to measure a whole lab, containing a human being (a friend of his), in any basis \cite{Wigner67}. Paradoxical conclusions may emerge from such situations due to the tension between the rules of evolution for isolated quantum systems, which in principle can be applied at any scale, and the need for the projection postulate to describe measurements performed by observers. While in most practical situations, it is clear whether or not an entity should be considered as a quantum system or an observer -- and therefore whether its interaction with the system should be described with a unitary evolution (Schr\"odinger equation) or via the projection postulate -- the transition between these two behaviors continues to cause much debate.  In particular it remains unclear where such a ``cut'' \cite{Heisenberg30} exists between systems that can be in quantum superposition and those which cannot either in principle or in practice.

In Ref.~\cite{Frauchiger18}, Frauchiger and Renner (FR) present an extended Wigner friend scenario involving two observers, the friends of Wigner, and two ``super-observers'', W and \textoverline{W}. 
The latter are assumed to be able to measure their friends and their labs in arbitrary bases of states.
FR's key result is to formulate apparently natural assumptions about this setup which they argue lead to an inconsistency. This study has triggered a large number of comments and articles re-examining the scenario. These new papers have identified hidden assumptions \cite{Sudbery19,Nurgalieva19}, gathered different arguments questioning FR's surprising conclusion \cite{Healey18,Lazarovici19,Fortin19,Waaijer19}, and generated new discussions of the quantum formalism and its interpretations \cite{Baumann18,Krismer18,Stacey19,Losada19,Nurgalieva20,Bub21}.

Below, we start from the assumption (made by Wigner and FR) that an observer can be described as a quantum system that becomes entangled with the system is measures. We then derive a necessary condition for such quantum system to behave as an observer, from the point of view of any other observer: the existence of a stable degree of freedom that we refer to as a \emph{memory}, which becomes entangled with the measured system and retains a record of the measurement outcome until the end of the experiment. 

We then use this result to introduce a simple reformulation of FR's scenario, based on an optical interferometer, which allows us to analyze the alleged paradox. Crucially, in our setup, only small systems that can be unambiguously treated quantum-mechanically are entangled. 
Although our reformulation arguably marks a philosophical departure from FR's discussion of a ``friend" reaching an entangled state, our model, in fact, leads to the same mathematical description. 
We then revisit the assumptions and properties used in Ref.~\cite{Frauchiger18} to formulate their paradox. We show that some of them require the memories to be erased by the super-observer, preventing the corresponding agents from behaving as observers. Conversely, others of these properties and assumptions require the memories to remain untouched, eliminating the main feature of the super-observer. In the end, the inconsistency only appears if one compares properties from two logically different contexts that we make explicit. We also demonstrate that the three properties involved in the paradox are associated with the values taken by observables which do not commute, and therefore are forbidden by quantum mechanics to take simultaneously well-defined values. Thus, we find that the argument of Ref.~\cite{Frauchiger18} does not lead to any inconsistency within quantum mechanics. 

By focusing on the role of memory, projection, and unitary evolution, we make explicit limits in the Wigner's Friend paradox and the mechanism of quantum measurement, and provide an operational prescription to identify devices behaving as an observer. This discussion is particularly useful in light of the recent related No-Go theorem of \cite{Bong20} that sought to illuminate key issues in quantum interpretation. We also note that the role of the material record of an observer's outcome has been examined \cite{Deutsch85} to introduce an experiment distinguishing between Everett's and the Copenhagen formulations of quantum mechanics, and recently to shed new light on the original Wigner's friend scenario \cite{Matzkin20}.

\begin{figure}[t]
\begin{center}
\includegraphics[width=0.5\textwidth]{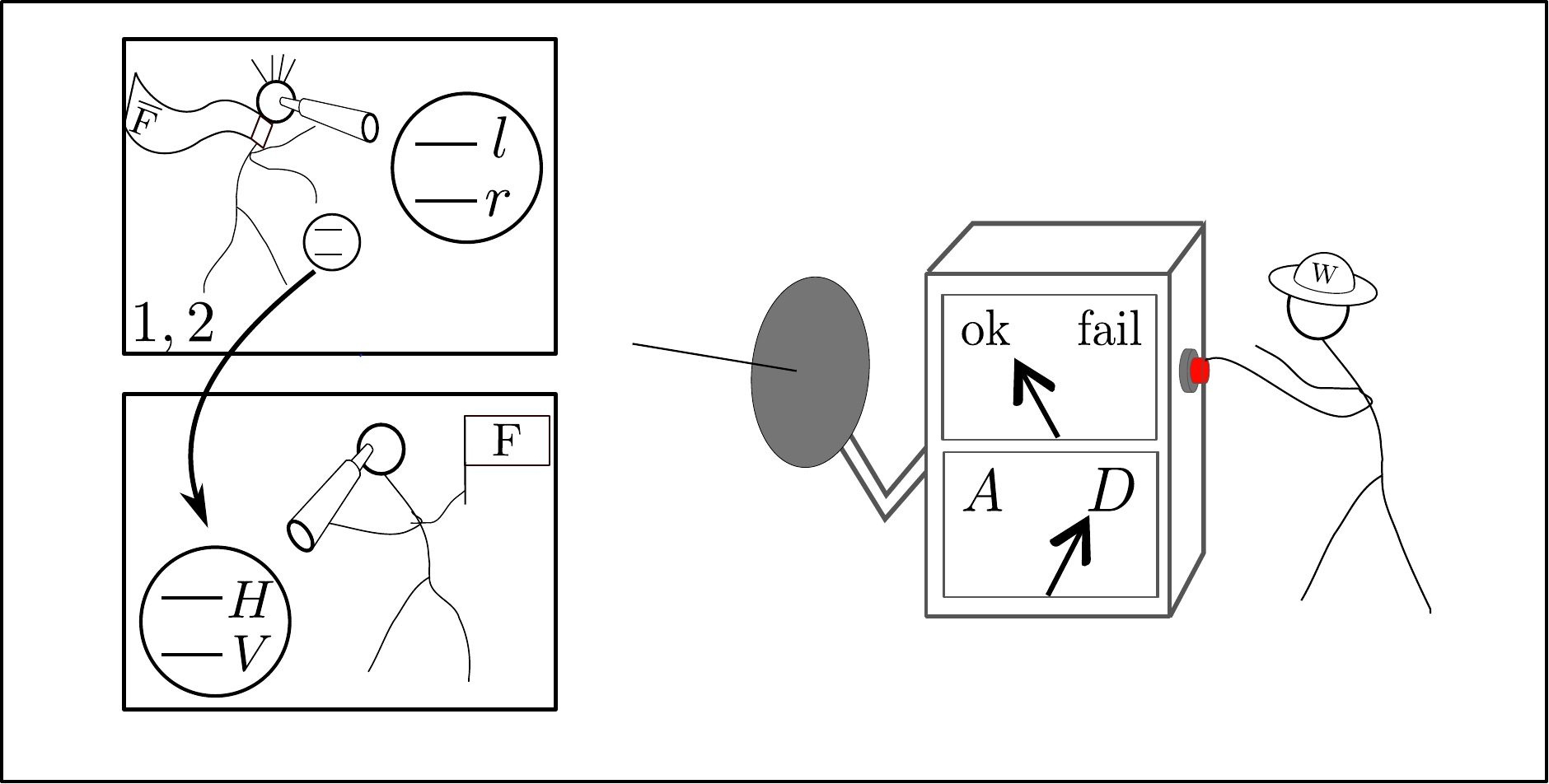}
\end{center}
\caption{Extended Wigner's friend scenario. Wigner's friend \textoverline{F}  (whose lab is described by states $1,2$) measures a first qubit in basis $\{l,r\}$, and depending on the outcome prepares another qubit and sends it to the F's lab. F measures the second qubit in basis $\{H,V\}$. Finally, Wigner measures both labs in the basis $\{(A,\text{ok}),(A,\text{fail}),(D,\text{ok}),(D,\text{fail})\}$.\label{f:setup}}
\end{figure}

\section{Results}

\subsection{Observers and memory} \label{s:Memory}

In most situations, it is easy to identify entities that behave as observers. The measurement an observer ${\cal O}$ performs on a quantum system is then well described by the measurement postulate, which attributes a well-defined outcome to the measurement and asserts that the measured system is projected into the corresponding eigenstate of the measured observable. The result of the measurement resides in the observer's memory. For specificity we can consider that ${\cal O}$ writes down the measurement results on a piece of paper, which also serves as a memory that the projection occurred. Another observer ${\cal O}'$ only has access to the result of the measurement either by asking ${\cal O}$, looking at the piece of paper, or immediately repeating ${\cal O}'s$ measurement. 

Wigner's friend and FR's paradoxes arise when assuming the existence of a super-observer. In contrast with a regular observer, a super-observer has full access to the whole quantum system comprising the memory of another observer and the measured system.  This larger quantum system is potentially a highly interacting collection of $>10^{26}$ atoms. Thus the meaning of ``full access" is that super-observers can manipulate all $>10^{26}$ atoms in their full entangled, superposed states to measure them in an arbitrary way. In particular, such measurement can create quantum superposition of macroscopically different states of the observer. As pointed out by Schr\"odinger and his famous cat gedankenexperiment \cite{Schrodinger35}, we never see macroscopic objects like a piece of paper with writing on it in a quantum superposition. Beyond the difficulty related to the size and complexity of such a large system itself, a pure superposition state is posited on the system being completely closed off from any ``environment'', which becomes increasingly difficult to achieve, in practice, as the system size grows. Apart from this practical limitation however, there is no evidence that closed systems should not be described by a unitary evolution (i.e. Schr\"odinger's equation) beyond a given size or complexity. This fact may seem in contradiction with the non-unitary projection predicted by the measurement postulate, and Wigner's friend gedankenexperiment (and extensions) precisely intend to investigate this tension. To do so, one generally assumes that (i) Wigner's technological means grants him such degree of control on another observer ${\cal O}$ which has performed a measurement of a quantum system $S$, and (ii) ${\cal O}$ and $S$ constitute a closed quantum system whose evolution is ruled by Schr\"odinger's equation. In this context, the measurement postulate is therefore assumed to be an effective evolution for the measured system, arising because one does not have access to the full state of ${\cal O}$ and $S$.

As Wigner's technology allows him to perform arbitrary measurements on the memory of ${\cal O}$, its complexity does not matter, and we can think of some properties of the measurement process with the following toy model: Let the system $S$ being measured be a qubit admitting the basis of states $\{\ket{0}_S,\ket{1}_S\}$ and initially described by state $\ket{\psi}_S = c_0\ket{0}_S+c_1\ket{1}_S$, and the memory of ${\cal O}$ be another qubit $M$ admitting the basis of states $\{\ket{0}_M,\ket{1}_M\}$. We assume that ${\cal O}$ measures the qubit in the basis $\{\ket{0}_S,\ket{1}_S\}$. The unitary evolution of the memory and the system prepares an entangled state:
\bb
\ket{\Psi}_{SM} = c_0\ket{0}_S\otimes\ket{0}_M+c_1\ket{1}_S\otimes\ket{1}_M.\label{eq:SMemEnt}
\ee
Such evolution can e.g.~be generated by the Schr\"odinger equation assuming a coupling Hamiltonian between $S$ and $M$ switched on only during a finite time. As long as the system $SM$ is in state $\ket{\Psi}_{SM}$, the memory qubit serves as a retrievable record that the system was found either in state $\ket{0}_S$ or $\ket{1}_S$. Such entangled state Eq.~\eqref{eq:SMemEnt} indeed captures the perfect correlation between the memory and system's state $\ket{0}_S$ and $\ket{0}_M$ on one hand, $\ket{1}_S$ and $\ket{1}_M$ on the other hand.  

To derive a necessary condition for such a unitary description of the memory and system evolution to faithfully model a quantum measurement, we now consider the point of view of a second independent observer ${\cal O}'$ measuring the qubit after ${\cal O}$. Contrary to Wigner, ${\cal O}'$ is a regular observer, who does not have access to the whole state of $S$ and the memory $M$ of ${\cal O}$. Therefore, the statistics of the measurement performed by ${\cal O}'$ should not depend on whether we described the measurement by ${\cal O}$ with the measurement postulate or with a unitary evolution, for the latter to be valid. One can indeed explicitly check that \emph{any} measurement on qubit $S$ yields the same statistics (see Appendix A) when computed from state $\ket{\Psi}_{SM}$ and from the state resulting of the application of the measurement postulate (assuming ${\cal O}'$ does not know the result of the measurement by ${\cal O}$):
\bb
\rho_S = \vert c_0\vert^2\ket{0}_S\bra{0} + \vert c_1\vert^2\ket{1}_S\bra{1} \label{eq:rhocollapsed}.
\ee
On the other hand, it may be enough to alter the state of $S$ and $M$ before the measurement of ${\cal O}'$, to cause the statistics of some measurements performed on qubit $S$ to differ from that obtained from Eq.~\eqref{eq:rhocollapsed} (see Appendix A for a direct proof). 
A striking situation corresponds to the case where one is able to totally erase the memory $M$, i.e. apply the inverse of the measurement unitary to prepare back the initial state $\ket{\psi}_S\ket{0}_M$. In this case, there is no evidence in any future operation done on the qubit that the measurement has ever been made. In other words, the unitary description of the measurement requires the memory to be stable (its joint state with the system must be unchanged) until the end of the experiment. One can make this condition more precise by considering that the memory is composed on many degrees of freedom (as the piece of paper certainly is) in which the measurement outcome is redundantly copied multiple times~\cite{Zurek09}. 
The memory-system state at the end of the interaction would then rather look like:
\bb
\ket{\Psi}_{SM_1....M_N} &=& c_0\ket{0}_S\otimes\ket{0}_{M_1}...\otimes\ket{0}_{M_N}\nonumber\\
&&+c_1\ket{1}_S\otimes\ket{1}_M...\otimes\ket{1}_{M_N}.\label{eq:SMemEntN}
\ee
In principle, it is then sufficient that \emph{one single} degree of freedom of the memory remains entangled with the measured qubit for the statistics of \emph{any} later measurement on $S$ to be the same as computed from state $\rho_S$. It is then clear that for a macroscopic memory, composed of many degrees of freedom able to become entangled with the system, it is practically impossible to perfectly inverse the unitary leading to Eq.~\eqref{eq:SMemEntN}, and evidence of the measurement will always remain in future measurement statistics. These observations allow us to formulate the following necessary condition:
\begin{condition}\label{c:Observer}
For an entity ${\cal O}$ to behave as an observer, there must exist at least one degree of freedom in which ${\cal O}$ can encode the results of measurement, and which is untouched for the duration of the considered experiment.
\end{condition}
This condition can also be obtained by considering that a defining property of a measurement is to generate an outcome. The Condition \ref{c:Observer} ensures that the operation performed by ${\cal O}$ results in registering an outcome that will survive subsequent stages of the experiment. In particular, to be used in Wigner's friend type scenarios, this condition applies to the friends solely if their memory remain stable even when Wigner makes his own measurements.

As a final remark for this section, we emphasize that entangled states as Eq.~\eqref{eq:SMemEnt} or  Eq.~\eqref{eq:SMemEntN} do not explain how a \emph{single} outcome is obtained from a readout of the memory. These states merely provides the probabilistic correlations that can be observed if such readout is made. The transition to a single definite outcome can only be described using the projection postulate to model the readout of the memory, as done for instance in von Neumann's seminal model of quantum measurement \cite{vonNeumann55}. This last projection can however be postponed to anytime after the interaction between the system and the memory without affecting the statistics of measurements performed by other observers. Meanwhile, the effect of all operations involving the memory (as in the scenario investigated below) can be taken into account.

\subsection{The Extended Wigner's friend scenario as an interferometer}

\subsubsection{Interferometric setup}

We now consider the following reformulation of the situation considered by FR in Ref.~\cite{Frauchiger18}. For a more detailed exact mapping of notations, see Table~\ref{t:Table}. The scenario involves two labs, each containing a qubit and an observer (referred to as a ``friend'' of Wigner), and two super-observers able to measure the two labs, W and \textoverline{W}. For the sake of simplicity, we merge W and \textoverline{W} into a single observer external to the labs, that we call Wigner.

The mere possibility of Wigner's full control over the labs prevents us from using the measurement postulate directly to describe the measurements done by his friends. We rather use the description presented in Section \ref{s:Memory} based on a unitary evolution of quantum systems playing the role of the friends' memories. On the other hand, as Wigner's measurement marks the end of the experiment, and it is assumed that nobody else will measure Wigner and his environment, we can describe Wigner's measurements via the usual projection postulate.

\begin{figure*}
\begin{center}
\includegraphics[width=0.98\textwidth]{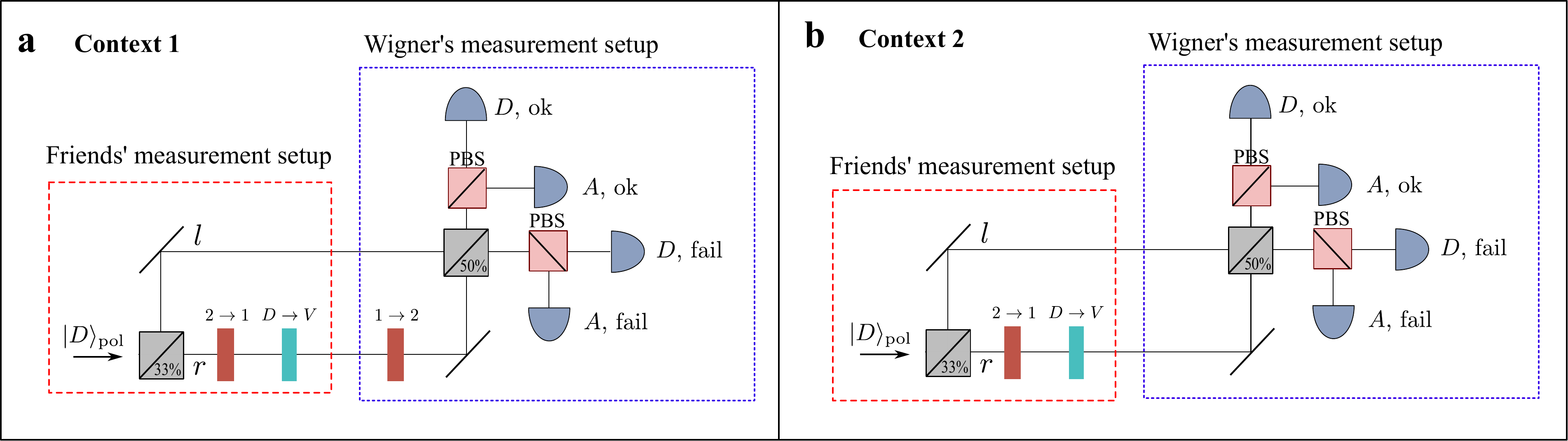}
\end{center}
\caption{\textbf{a}: Proposed inteferometric setup to implement the extended Wigner's friend scenario, when Wigner is indeed a super-observer (context 1). The photon is initially prepared in polarization state $\ket{D}_\text{pol}$ and a wavepacket of shape $s=2$. The red box corresponds to the two friends' measurements. The first beam-splitter is assumed to have $1/3$ transmission probability, such that after the first beam-splitter of the interferometer (which does not to affect the wave-form), it is described by the state $\ket{\Psi_0} = \frac{1}{\sqrt{3}}(\ket{D,r,2}+\sqrt{2}\ket{D,l,2})$. \textoverline{F} measures the which-path information. In our setup, this corresponds to a correlation between the mode shape of the photon and the path, that can be generated e.g. if an optical element causing the transition from state $\ket{2}_\text{shape}$ to the orthogonal state $\ket{1}_\text{shape}$, is inserted in the arm $r$, yielding the photon state $\ket{\Psi_1} = \frac{1}{\sqrt{3}}(\ket{D,r,\red{1}}+\sqrt{2}\ket{D,l,2}).$ Then, in the scenario of Ref~\cite{Frauchiger18}, \textoverline{F} prepares the state of the other qubit, encoded in the polarization, depending on its outcome. This step does not really requires \textoverline{F} to read out the memory, and can be rather done by correlating directly the memory (the shape of the photon) with the polarization. This can be achieved assuming the presence of a chiral crystal (or waveplate) in path $r$ rotating the polarization of the photon by $45\ensuremath{^\circ}$, transforming the diagonal polarization state into the vertical polarization. This yields state $\ket{\Psi_2}$ expressed in Eq.~\eqref{eq:psi2}. Finally, Wigner performs a measurement in the basis $\{(D,ok), (A,ok), (D,fail), (A,fail)\}$. In our setup, this can be achieved using the elements gathered in the blue box, namely, a mode-shaper turning the photon wavepacket in arm $r$ from state $s=1$ to $s=2$, a balanced beam-splitter (which acts only on the path degree of freedom), two polarized beam-splitters (PBS) which transmit diagonally-polarized photons and reflect antidiagonally-polarized ones and four photon counters. 
\textbf{b}: Setup corresponding to context 2, where Wigner is not a super-observer. The difference is the absence of the second mode-shaper inside the blue box. \label{f:contexts}}
\end{figure*}


In our formulation, the role of the two friends' memories, and the qubits they measure, are all played by different degrees of freedom of a single photon traveling through a Mach-Zender interferometer (see Fig.~\ref{f:setup}). These degrees of freedom can all, for our purpose, be modeled by qubits (three in total).
First, the path taken by the photon (the arm of the interferometer it travels through) plays the role of one of the two qubits. This qubit initially belongs to friend \textoverline{F} who is able to measure it. 
The memory of \textoverline{F} is represented by the spatiotemporal shape of the photon wavepacket (or equivalently the mode of the waveguide the photon is in).
Finally, the polarization of the photon plays the role of the other qubit and the memory of the other friend F of Wigner, all together. While one could think of a more sophisticated version of the setup allowing us to distinguish these roles, it will be enough for our purpose to work with these three qubits. We can specify the photon state using the orthogonal basis $\{\ket{\alpha,n,s}\}$. The states $\ket{\alpha,n,s} \equiv \ket{\alpha}_\text{pol}\otimes\ket{n}_\text{path}\otimes\ket{s}_\text{shape}$ are labeled by the polarization $\alpha = H,V$, the path taken by the photon $n =l,r$ and the two possible orthogonal shapes of its wavepacket $s = 1,2$.

\begin{figure}
\begin{center}
\includegraphics[width=0.49\textwidth]{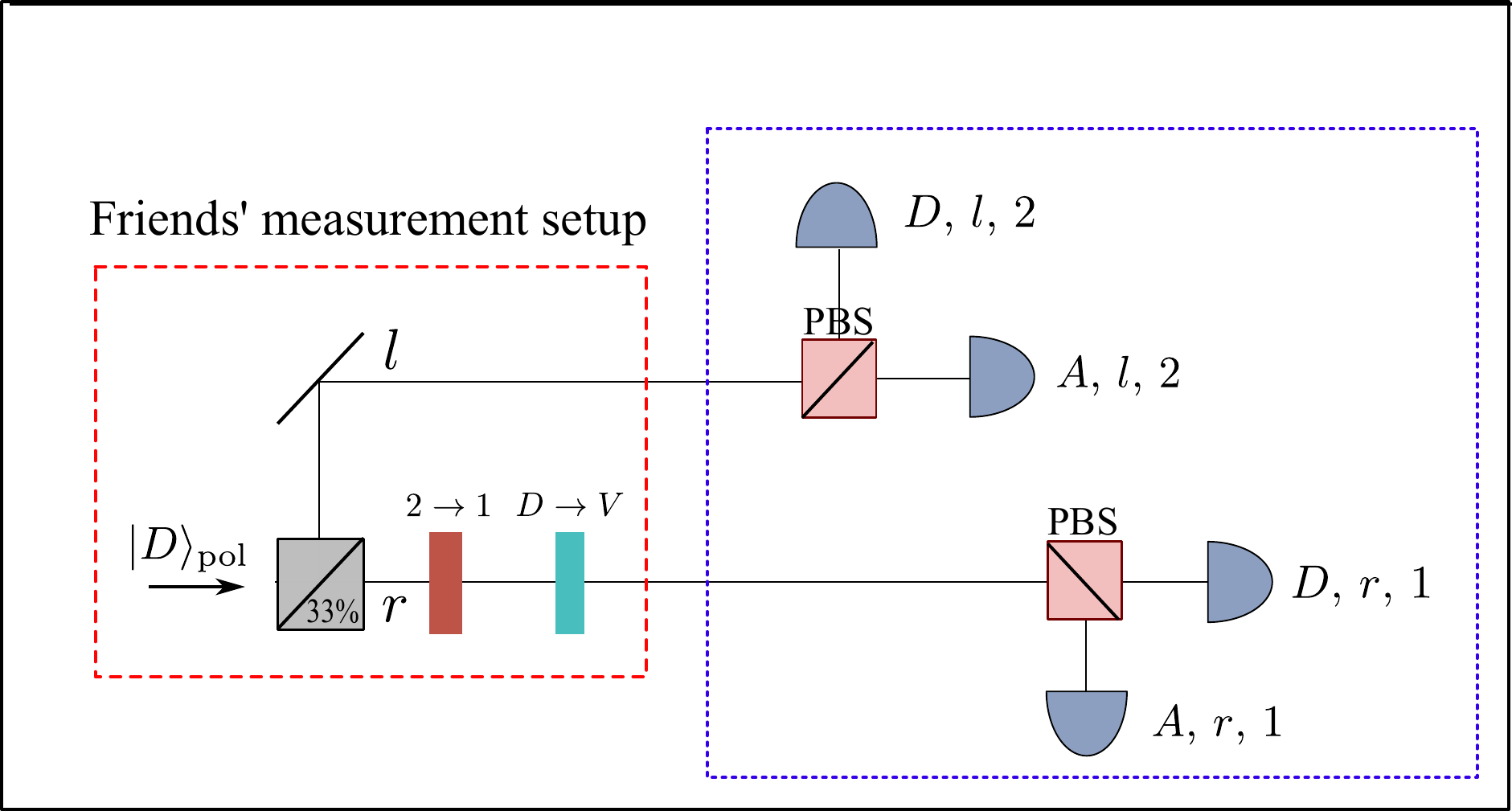}
\end{center}
\caption{Measurement setup allowing Wigner to test the correlations in state $\ket{\Psi_2}$ right after the interaction with the friends. \label{f:correl}}
\end{figure}

The unitary evolutions associated with the two friends' measurements can be implemented owing to the optical elements in the red box of Fig.~\ref{f:contexts}\textbf{a}, preparing state:
\bb
\ket{\Psi_2} = \frac{1}{\sqrt{3}}\Big(\ket{V,r,1}+\sqrt{2}\ket{D,l,2}\Big).\label{eq:psi2}
\ee
We have introduced the diagonal $D$ and antidiagonal $A$ polarization states
\begin{subequations}
\bb
\ket{D}_\text{pol} &=& \tfrac{1}{\sqrt 2}\Big(\ket{H}_\text{pol}+\ket{V}_\text{pol}\Big),\\
\ket{A}_\text{pol} &=& \tfrac{1}{\sqrt 2}\Big(\ket{H}_\text{pol}-\ket{V}_\text{pol}\Big).
\ee 
\end{subequations}
%
$\ket{\Psi_2}$ describes the photon exiting the friends' lab. At this point, Wigner could use photon-counters to measure whether the photon took arm $r$ or $l$. Furthermore, he could use a polarizing beam-splitter placed before the photodetectors to test the correlations between the polarization and the path, as shown in Fig.~\ref{f:correl}. Eq.~\eqref{eq:psi2} implies that Wigner would find perfect correlations between the photon taking arm $r$ and having polarization $V$, and between the photon taking arm $l$ and having polarization $D$, which can be summarized in the properties.


\begin{property}\label{p:HI}
The probability for the photon to travel through arm $r$ (and therefore having shape $1$) and having polarization H is zero,
\end{property}
and 
\begin{property}\label{p:AII}
The probability for the photon to travel through arm $l$ (and therefore having shape $2$) and having polarization A is zero.
\end{property}

{\renewcommand{\arraystretch}{2.5}
\begin{table*}\begin{center}
\begin{tabularx}{18cm}{|l||Y|Y|}
\hline
& Present notations & Frauchiger-Renner article\\
\hhline{|=::=:=|}
First qubit & Path of the photon & Quantum coin R\\
\hline
Measurement basis of \textoverline{F}  & $\{\ket{r}_\text{path},\ket{l}_\text{path}\}$ & $\{\ket{\text{tails}}_\text{R},\ket{\text{heads}}_\text{R}\}$\\
\hline
Lab states of \textoverline{F} & $\{\ket{l,1},\ket{r,2},...\}$  & $\{\ket{{\mathrm{\overline t}}}_{\mathrm{\overline L}},\ket{{\mathrm{\overline h}}}_{\mathrm{\overline L}},...\}$ \\
\hline
Basis of Wigner's measurement on \textoverline{F} & $\{\ket{\text{ok}},\ket{\text{fail}}\}$ where $\ket{\text{fail}} = \tfrac{1}{\sqrt 2}\Big(\ket{r,1}+\ket{l,2}\Big)$ & $\{\ket{\mathrm{\overline{ok}}}_\mathrm{\overline{L}},\ket{\mathrm{\overline{fail}}}_\mathrm{\overline{L}}\}$ where $\ket{\mathrm{\overline{fail}}}_\mathrm{\overline{L}} = \tfrac{1}{\sqrt 2}\Big(\ket{\mathrm{\overline t}}_{\mathrm{\overline L}}+\ket{\mathrm{\overline h}}_{\mathrm{\overline L}}\Big)$
\\
\hhline{|=::=:=|}
Second qubit & Polarization & Spin $1/2$ S\\
\hline
Measurement basis of F & $\{\ket{H}_\text{pol},\ket{V}_\text{pol}\}$ & $\{\ket{\uparrow}_\text{S},\ket{\downarrow}_\text{S}\}$\\
\hline
Lab states of F & $\{\ket{H}_\text{pol},\ket{V}_\text{pol}\}$ & $\{\ket{\tfrac{1}{2}}_\text{L},\ket{-\tfrac{1}{2}}_\text{L}\}$\\
\hline
Basis of Wigner's measurement on F & $\{\ket{D}_\text{pol},\ket{A}_\text{pol}\}$ where $\ket{D}_\text{pol} = \tfrac{1}{\sqrt 2}\Big(\ket{H}_\text{pol}+\ket{V}_\text{pol}\Big)$ & $\{\ket{\mathrm{{ok}}}_\mathrm{{L}},\ket{\mathrm{{fail}}}_\mathrm{{L}}\}$ \\
\hline
\hline
\end{tabularx}
\end{center}
\caption{Correspondance of notations between our setup and Ref.~\cite{Frauchiger18}.}\label{t:Table}
\end{table*}
}

\subsubsection{Wigner's ``super-measurements''}
\label{s:WignerSuperMeas}

The end of the protocol corresponds to Wigner's measurements. Rather than doing the measurement described above, made to check  properties 1 and 2, Wigner wants to 
probe bases different from those in which the friends did their own measurements. 
The first qubit was measured in the $l,r$ basis, and the outcome was copied in the shape 1,2. Wigner decides to measure the joint qubit-lab state in a basis containing the states $\{\ket{\text{ok}},\ket{\text{fail}}\}$, with 
\begin{subequations}
\bb
\ket{\text{fail}} &=& \frac{1}{\sqrt 2}(\ket{r}_\text{path}\otimes\ket{1}_\text{shape}+\ket{l}_\text{path}\otimes\ket{2}_\text{shape}),\quad\quad\\
\ket{\text{ok}} &=& \frac{1}{\sqrt 2}(\ket{r}_\text{path}\otimes\ket{1}_\text{shape}-\ket{l}_\text{path}\otimes\ket{2}_\text{shape}).\quad\quad
\ee
\end{subequations}

The second memory-qubit state is encoded in the polarization that 
Wigner chooses to measure in the basis $\{ \ket{D}_\text{pol},\ket{A}_\text{pol}\}$. 
It is useful to express the photon state in the $\{\text{ok},\text{fail}\}$ basis:
\begin{widetext}
\begin{subequations}\label{Psi}
\bb
\ket{\Psi} &=& \frac{1}{\sqrt{6}}\Big(\blue{\overbrace{\ket{V,\text{fail}}+\ket{V,\text{ok}}}^r}+\red{\overbrace{\ket{V,\text{fail}}-\ket{V,\text{ok}}+\ket{H,\text{fail}}-\ket{H,\text{ok}}}^{l}}\Big)\label{Psia}\\
&=& \frac{1}{\sqrt{12}}\Big(3\ket{D,\text{fail}}-\ket{D,\text{ok}}-\ket{A,\text{fail}}-\ket{A,\text{ok}}\Big).\label{Psic}
\ee
\end{subequations}
\end{widetext}
From the state above, we can compute the probability of all four outcomes. In particular, the probability of finding (A,ok) is $\vert \bra{A,\text{ok}}\Psi\rangle\vert^2 = 1/12$.

In Eq.~\eqref{Psia}, we have indicated in blue (resp. red) the terms coming from the photon traveling through arm $r$ (resp. $l$) for later discussion. We see that the interference between the two arms is responsible for the cancellation of the terms proportional to $\ket{V,\text{ok}}$. It is important for later to record the following property:

\begin{property}\label{p:Vok}
Due to the interference between the two photon paths, the amplitude of the photon reaching one of the ports labeled by ``ok''  and having polarization V is zero.\\
\end{property}


\subsection{The paradox}

In Ref.~\cite{Frauchiger18}, FR use a set of assumptions to point out three properties of the measurement outcome statistics, that they claim are paradoxical. Their intention is to show that no physical theory can satisfy these three assumptions simultaneously. In our words and notations, the assumptions are:

\begin{itemize}
\item (U): \emph{The measurements performed by the two friends of Wigner can be described by Wigner as a unitary (entangling) evolution of the qubits and the friend's labs}. This is not written explicitly in Ref.~\cite{Frauchiger18}, but is used in their analysis, as pointed out by some comments \cite{Nurgalieva19,Sudbery19}. It corresponds to describing the friends' measurements as interaction with memories rather than using the projection postulate, as we did above. This assumption requires than the memory and the measured system form a closed system. It also implies than the memory can be manipulated after the interaction with the system, and that in principle future evidence of the measurement can be erased.
\item (Q): \emph{If a quantum system is in a state orthogonal to one of the eigenstates of an observable, then a measurement of this observable has zero probability to yield the corresponding outcome}. This assumption is formulated differently in Ref.~\cite{Frauchiger18}, but is used with this meaning in their analysis, as pointed out by some comments \cite{Sudbery19}. This assumption is included in Born's rule to compute measurement statistics.
\item (S): \emph{For any measurement performed by a given observer, a single definite outcome is obtained}. Note that this assumption requires that at the end of the experiment, the memory is read out.
\item (C): \emph{There is consistency between expectations for measurement outcomes predicted based on the outcomes of different observers, even when one observer is actually able to measure another one (or another's one memory)}. This can be stated simply as: Different observers should not find different results for observations on the same system. Note that FR's argument solely requires consistency between outcomes that are certain (to which Born's rule assign a unit probability).
\end{itemize}

From these assumptions, FR deduce in Ref.~\cite{Frauchiger18} that their setup should verify Properties \ref{p:HI}, \ref{p:AII} and \ref{p:Vok} at the same time, and that the measurement statistics should be captured by state $\ket{\Psi}$. However, these three properties combined seem to rule out the possibility to obtain the outcome $(A,ok)$, according to the following reasoning: 
\begin{enumerate}
\item Property \ref{p:AII} forbids the outcome $A$ to be obtained when the first qubit is found in state $\ket{l}_\text{path}$, which means that $(A,\text{ok})$ is only compatible with the first qubit being found in $\ket{r}_\text{path}$.\\
\item Then Property \ref{p:HI} implies that when the first qubit is
found in $\ket{r}_\text{path}$, the photon must be in polarization state $V$.\\
\item Finally, Property \ref{p:Vok} can be used to state that if the photon is in state $V$, it cannot be in state $\ket{\text{ok}}$, such that finally $(A,\text{ok})$ should be forbidden.
\end{enumerate}
This conclusion is paradoxical because the statistics of outcomes computed from state $\ket{\Psi}$ predicts a non-zero probability of obtaining the outcome $(A,\text{ok})$\footnote{In FR's article, one considers that the experiment is repeated multiple time until outcome $(A,\text{ok})$ is obtained, which corresponds to an occurrence of the paradox.}.

\subsection{Insights from the interferometric setup}

Discussions of this paradox have often involved questioning the assumptions made in Ref.~\cite{Frauchiger18}, as in e.g. Refs.~\cite{Healey18,Lazarovici19,Sudbery19,Nurgalieva19}. Several arguments have been used to show that quantum mechanical setups do not verify all of them. For instance, it is stated in Ref.~\cite{Healey18} that the violation of a special variation of Bell inequalities, whose experimental verification is reported in Ref.~\cite{Proietti19}, rules out assumption (C).\\

Here we take another approach by studying how the paradox would arise in a realistic setup, involving systems whose dynamics are known to obey to quantum mechanics. As an advantage, we therefore do not have to make assumptions (U), (Q) (S), (C) to predict the outcomes of measurements made on the system. As in the case of FR, we find that the statistics of Wigner's measurement outcomes are given by state $\ket{\Psi}$. Meanwhile, we expressed the three paradoxical properties \ref{p:HI}, \ref{p:AII} and \ref{p:Vok} as a function of the interferometer's degrees of freedom. 

The latter connection allows us to stress a crucial point concerning properties \ref{p:HI} to \ref{p:Vok}, which is that they belong to two different \emph{contexts}, i.e.~two different incompatible choices of experimental setups. Indeed, Properties \ref{p:HI}, \ref{p:AII} refer to the which-path information $\{r,l\}$, while Property \ref{p:Vok} refers to a different basis involving coherent superpositions of the path states $\{\ket{l},\ket{r}\}$. In the absence of the second beam-splitter closing the interferometer, the path taken by the photon can be measured (see Fig.~\ref{f:correl}): a click at one of the photon detectors causes the photon to take a definite path $\ket{l}$ or $\ket{r}$, and the validity of Properties \ref{p:HI}, \ref{p:AII} can be checked. Conversely, when the beam-splitter is present, it ensures that the which-path information remains unavailable, i.e~\emph{does not take any definite value} after a click at any of the detectors. As known from single-photon interferometer experiments, this unvailability of the which-path information is necessary for the interference between the two paths to take place, which in turn is needed for Property \ref{p:Vok} to hold (see Section \ref{s:WignerSuperMeas}), but the validity of properties \ref{p:HI} and \ref{p:AII} cannot be checked.

Remembering that the shape corresponds to the memory in which \textoverline{F} registers the outcome of the which-path measurement, one can also give the following interpretation: the mode-shaper required to implement Wigner's measurement unitarily \emph{erases} the memory of \textoverline{F}. As explained in Section \ref{s:Memory}, this means that Condition \ref{c:Observer} is violated for \textoverline{F} and therefore no trace that the which-path measurement has been made remains.

\subsection{Varying the context}

We can expand this discussion by considering variations on the setting just described, in a way that ensures the validity of properties \ref{p:HI} and \ref{p:AII} (and that \textoverline{F} behaves as an observer). Referring to the ``super-measurement'' by Wigner we have described in previous section as Context 1, we instead consider a contrasting scenario in which Wigner performs a regular measurement, leaving some degrees of freedom in the memory untouched. Without loss of generality, we identify these degrees of freedom with the photon wavefunction shape $s$. This new situation corresponds to a modified measurement setup (see Fig.~\ref{f:setup}\textbf{b}) that we call Context 2 whose net effect is a simultaneous measurement of the polarization in the $D,A$ basis, and of the path in basis:
\begin{subequations}
\bb
\ket{\text{fail}'}_\text{path} &=& \frac{1}{\sqrt 2}(\ket{r}_\text{path}+\ket{l}_\text{path}),\quad\quad\\
\ket{\text{ok}'}_\text{path} &=& \frac{1}{\sqrt 2}(\ket{r}_\text{path}-\ket{l}_\text{path}),\quad\quad
\ee
\end{subequations}
%
In the new measurement basis, the state $\ket{\Psi_2}$ reads:

\begin{widetext}
\begin{subequations}\label{Psip}
\bb
\ket{\Psi'} &=& \frac{1}{\sqrt{6}}\Big(\blue{\overbrace{\ket{V,\text{fail}',1}+\ket{V,\text{ok}',1}}^r}+\red{\overbrace{\ket{V,\text{fail}',2}-\ket{V,\text{ok}',2}+\ket{H,\text{fail}',2}-\ket{H,\text{ok}',2}}^{l}}\Big)\label{Psipa}\\
&=& \frac{1}{\sqrt{12}}\Big(\ket{D,\text{fail}'}\otimes(\ket{1}_\text{shape}+2\ket{2}_\text{shape})-\ket{D,\text{ok}'}\otimes(\ket{1}_\text{shape}-2\ket{2}_\text{shape})\nonumber\\
&&\quad-\ket{A,\text{fail}'}\otimes\ket{1}_\text{shape}-\ket{A,\text{ok}'}\otimes(\ket{1}_\text{shape}-2\ket{2}_\text{shape})\Big).\label{Psipc}
\ee
\end{subequations}
\end{widetext}
When the setting of Fig.~\ref{f:setup}\textbf{b} is used, the information about the path taken by the photon is preserved in the shape and the validity of Properties \ref{p:HI} and \ref{p:AII} can be checked. However, we can see that the interference between the two paths \emph{does not occur anymore} and the probability of outcome $\ket{V,\text{ok}'}$ does not vanish
$$\| \tfrac{1}{\sqrt 6}(\ket{V,\text{ok}',1}-\ket{V,\text{ok}',2})\|^2 = 1/3,$$
which means that Property \ref{p:Vok} is automatically violated. 
Therefore, neither of these two contexts allow all of properties 1 through 3 to hold simultaneously.

One may consider whether another measurement setup, that does allow all three properties to remain simultaneously valid, exists.
The answer is no, and this can be proven by noting that Properties \ref{p:HI} to \ref{p:Vok} correspond to assertions on the values taken by a set of \emph{non-commuting} observables, which are consequently forbidden by quantum mechanics to simultaneously all accept a well-defined value (see Appendix B)\footnote{In the language of quantum contextuality, one can say that Properties 1, 2 and 3 are associated with the rays defined by $P_1 = \ket{H,r}\bra{H,r}$, $P_2 =\ket{A,l}\bra{A,l}$ and $P_3 =\ket{V,\text{ok}}\bra{V,\text{ok}}$, respectively, taking the value $0$. However, ray $P_1$ can be jointly measured with $P_2$ or with $P_3$ (in the sense that one can form sets of orthogonal projectors summing to one which include both $P_1$ and $P_2$ or $P_1$ and $P_3$, but not with both simultaneously.}.

\subsection{The interferometer and the assumptions of Frauchiger and Renner}

Eventually, it is enlightening to look at how the present setup fits within the assumptions of FR. We first note that for this setup Assumption (Q) is fulfilled as the system under investigation is clearly an isolated quantum system. 
We also presented the rationale for using Assumption (U) in \ref{s:Memory}.
Interestingly, the fate of the two other assumptions depends on whether Wigner is chosen to be a ``super-observer'' (Context 1 leading to state $\ket{\Psi}$) or a regular observer (Context 2 leading to state $\ket{\Psi'}$), i.e.~by inserting, or not, the second mode-shaper in arm $r$.

If Wigner has full control over his friends' labs (second mode-shaper present), his measurement statistics are captured by state $\ket{\Psi}$. However, one can see explicitly that assumption (S) is violated. Indeed, the memory of \textoverline{F}'s lab is erased during Wigner's measurement process before being read, and the branch of the wavefunction $\ket{\Psi_2}$ corresponding to \textoverline{F} finding path $l$ interferes with that corresponding to finding path $r$. The situation is reminiscent of the ``quantum eraser'' where a measurement is used to erase the photon which-path information after it was recorded, thereby restoring interference \cite{Scully82,Kim00,Walborn02}. As said above, the fact that the which-path information is made unavailable is necessary to the validity of property \ref{p:Vok}.
Consequently, the first friend does not have a definite measurement outcome (which would be either $r$ or $l$), which contradicts assumption (S). Note that the validity of (C) is hard to analyze in this case, as pretending we could take the point of view of the friend, who is in a superposition state, is hazardous. The consequence is that properties \ref{p:HI} and \ref{p:AII} do not hold anymore, such that the paradox does not arise. 

Conversely, if we assume that there exists some preserved record of the which-path information (second mode-shaper absent), assumptions (S) and (C) are verified. However, the state $\ket{\Psi'}$ is now the proper description for the outcome statistics, and Property \ref{p:Vok} does not hold anymore. Once again, the paradox does not arise. In summary, the paradox appears when comparing properties associated to two different contexts, i.e.~two different experimental setups shown in Fig.~\ref{f:setup}.

Finally, our results can be connected to a recently introduced No-go theorem Ref.~\cite{Bong20}, invalidating assumptions (S) and (C) in the presence of a super-observer, when assuming observers' freedom of choice and locality of measurements.

\section{Discussion}

We have identified a condition for a quantum system to behave as an observer, which is to possess a stable memory. Then, we have reformulated an extended Wigner's friend scenario introduced in Ref.~\cite{Frauchiger18} as an interferometric setup, involving three different degrees of freedom of a single photon to play the roles of two qubits and the memory of the friends measuring them. By analyzing this setup, we have shown that the three properties highlighted to be paradoxical correspond to two different contexts, in which the which-path information takes well defined values or not. These two contexts correspond to two different measurement setups, which access the values of different sets of observables, which we show do not commute. They are then forbidden by quantum mechanics to all simultaneously take well-defined values. As a consequence, the paradox never arises in any physical setup obeying quantum mechanics.

The fact that the three properties considered to formulate the paradox cannot hold simultaneously was already argued in Ref.~\cite{Lazarovici19}. We here illustrate the transition between the validity regime of Properties \ref{p:HI} and \ref{p:AII} and that of Property \ref{p:Vok}. In doing so, we can specifically identify the transition between the coherent (quantum) and incoherent (observer) behaviors of the friends. This transition is dictated by the presence, or not, of an untouched memory degree of freedom. We also note that our interferometric setup has similarities with Hardy's paradox \cite{Hardy92}, with an additionnal degree of freedom (the shape of the photon) to control the interference.

Our work is relevant however to more than just the FR thought experiment. Our results speak to unresolved issues in the Wigner's Friend paradox by identifying the state specific role a ``memory'' must play for an observer in the paradox. Thus, the transition between the two experimental contexts (captured by states $\ket{\Psi}$ and $\ket{\Psi'}$ respectively) can be interpreted as an assumption about the location of Heisenberg's cut \cite{Heisenberg30}. More precisely, it allows us to understand which apparatuses must be considered as observers -- the `ultimate measuring instruments' of Bohr \cite{Bohr39}, which verify Condition \ref{c:Observer} -- or included in the system described by quantum-mechanics \cite{Bub21,Bub19}. 

It may seem from Condition~\ref{c:Observer} that the status of a device (observer or not) depends on future choices made by super-observers. One can actually distinguish two cases: First, simple enough systems, such as single photons, over which one can have a large degree of control and deliberately choose to alter or not the state. In this case, it is indeed a \emph{choice} to have them behave as observers, and this choice can be delayed until after they interacted with the system. One could argue that rather than being observers, those systems \emph{can behave as such} in specific situations. On the other hand, it is easy to identify a class of systems complex enough that when they are used as memories, the unitary control required to erase their content is definitely out of reach. Those systems satisfy Condition~\ref{c:Observer} at any time and can be qualified as observer without ambiguity. 
In this discussion, the recent technological advances allowing the unitary manipulation of larger and larger quantum systems can be seen as a (slight) shift of which systems belong to this class. An important point is that for any (a priori given) experimental resolution, one can find systems belonging to this class, which can therefore be defined as observers' memory, and their action modeled through the measurement postulate. From their mathematical treatment of the friends and qubits' state, it is clear than FR make the assumption of technological means far beyond current technology level, such that measurements can be made in arbitrary bases even on systems such as human beings. In our prescription, making this assumption implies that, just as single photons, the ability of Wigner's friend to behave as an observer is not granted, but conditioned to the fact their memory is not erased during the experiment we consider. While this holds true in FR's scenario for the two agents W and \textoverline{W} (that we have merged into Wigner), this condition is clearly violated for F and \textoverline{F} which are therefore forbidden to behave as observers by construction.

More broadly, interpretations of quantum mechanics treating observers and quantum systems on different footing are often criticized because of the apparent flexibility in the position where such this cut should be placed. However, our setup, and the notion of memory, can be used to argue that the place of this cut is actually imposed by the practical (and objective) resolution of the experimentalist's apparatuses. 
Because different experimentalists with different apparatuses may be able to control different degrees of freedom, this approach can also be naturally related to a ``QBist'' interpretation \cite{Bub07,Fuchs09,Fuchs16} where the quantum state is ascribed by a given observer to a quantum system and may therefore be observer-dependent. A second point of contact with QBist approaches come with the emphasis on memory and its quantum accessibility. QBism holds that quantum states are epistemological rather than ontological. In particular, given the fundamental nature of probabilities in quantum states, QBism stresses that they should be seen as bets, conditioned by priors, placed on the results of an experiment. The priors are then updated once the experimental results are obtained.  Priors are, essentially, memories.  They are information about previous states of the world held by an observer and used to calculate quantum states.  Thus by showing how memory, as a manipulable quantum system, must function in Wigner's Friend argument, our setup may help articulate the ways in which quantum states in general should be viewed. Further comments regarding a QBist treatment of the problem can be found in Refs.~\cite{Stacey19,DeBrota20}.

\section{Acknowledgements}
We thank Joseph Eberly, Zachery Brown, \'Etienne Jussiau, Tathagata Karmakar, Joseph Murphree, John Steinmetz and Jing Yang for stimulating discussions that led to this paper. We acknowledge funding from NSF grant no. DMR-1809343 and John Templeton Foundation, award number 61835.

\bibliographystyle{plain}

\onecolumn\newpage
\appendix


\section{Appendix: Equivalence of measurement descriptions}

The equivalence of the entangled state Eq.~\eqref{eq:SMemEnt} and the collapsed state Eq.~\eqref{eq:rhocollapsed} for subsequent measurements can be showed by computing the probability to obtain an eigenstate $\ket{a}_S$ of an arbitrary observable $\hat A_S$ of qubit $S$:
\bb
{}_{SM}\bra{\Psi} a\rangle_S\langle a \ket{\Psi}_{SM}  &=&  \vert c_0\vert^2\vert{}_S\langle a \ket{0}_S\vert^2 + \vert c_1\vert^2\vert {}_S\bra{a} 1\rangle_S\vert^2\nonumber\\
&=& {}_S\bra{a}\rho_S \ket{a}_S.\label{eq:stat}
\ee

This implies that the statistics of subsequent measurement on the qubit $S$ is captured identically by both states. This property is a direct consequence of the perfect correlation of the memory states $\ket{0}_M$ and $\ket{0}_M$ with the system states $\ket{0}_S$ and $\ket{0}_S$, respectively. A unitary operation on the memory and the system will in general generate a state $\ket{\Psi^\text{gen}}_{SM} = \sum_{i,j=0,1} d_{ij}\ket{i}_S\otimes \ket{j}_M$ that will yield probability:
\bb
{}_{SM}\bra{\Psi^\text{gen}} a\rangle_S\langle a \ket{\Psi^\text{gen}}_{SM}  &=&  \sum_{i,j,k=0,1}d_{i,j}^*d_{k,j}{}_S\bra a \rangle_S\langle a \ket{j}_S\label{eq:statgen}\nonumber\\
\ee
of finding eigenstate $\vert  a \rangle_S$. It is now easy to check that only the states $\ket{\Psi^\text{gen}}_{SM}$ which verify the specific conditions (1) $\sum_{j=0,1} \vert d_{ij}\vert^2 = \vert c_{i}\vert^2$ and (2) $d_{00}^* d_{10} + d^*_{01} d_{11} = 0$ preserve the probability distribution \eqref{eq:stat} for any $\ket{a}_S$. This corresponds to unitary transformations which acts on the memory state and performs a basis rotation, transforming $\ket{0}_M$ and $\ket{1}_M$ to two orthogonal states. Any unitary transformation that does not respect this specific property, for instance that acts on the memory in a way conditionned to the system's state, will change the measurement statistics. In particular, the unitary transformation involved in Wigner's ``super-measurement'' (see Fig.~\ref{f:setup}\textbf{b}) erases the memory, yielding a state of the form $\ket{\Psi^\text{er}}_{SM} = (c_0\ket{0}_S+c_1\ket{0}_S)\otimes \ket{0}_M$ which implies a probability ${}_{SM}\bra{\Psi^\text{er}} a\rangle_S\langle a \ket{\Psi^\text{er}}_{SM} = \sum_{i,j} c^*_i c_j {}_S\bra{i}  a \rangle_S\langle a \ket{j}_S$ which is not equal to ${}_S\bra{a}\rho_S \ket{a}_S$ as long as $c_i$ and $c_j$ both differ from $0$.

\section{Appendix: Paradoxical properties as incompatible observables}

We introduce the following photon observables:
\bb
{\cal O}_1 &=& \ket{H,r,1}\bra{H,r,1}\nonumber\\
{\cal O}_2 &=& \ket{A,l,2}\bra{A,l,2}\nonumber\\
{\cal O}_3 &=& \ket{V,\text{ok}}\bra{V,\text{ok}}.
\ee
These three-qubit observables involve all three degrees of freedom of the photon. They are projectors admitting eigenvalues $0$ and $1$. It is easy to check that:
\begin{align}
[{\cal O}_1,{\cal O}_2] &= [{\cal O}_1,{\cal O}_3]=0,\nonumber\\
[{\cal O}_2,{\cal O}_3] &= \tfrac{1}{2}(\ket{A,l,2}\bra{V,\text{ok}}+\ket{V,\text{ok}}\bra{A,l,2}) \neq 0.
\end{align}
Meanwhile, saying properties \ref{p:HI}, \ref{p:AII} and \ref{p:Vok} hold is equivalent to say that, respectively, ${\cal O}_1$, ${\cal O}_2$ and ${\cal O}_3$ takes the value $0$. The fact that ${\cal O}_2$ and ${\cal O}_3$ do not commute then rules out the possibility for these three properties to hold simultaneously in any measurement setup.

\end{document}